\documentclass[final,numberedheadings]{aipproc}

\layoutstyle{6x9}

\newcommand{\rs}{\langle r^2\rangle\rule[-0.2em]{0em}{0em}_s}
\newcommand{\lbar}{\bar{\ell}}
\newcommand{\Pint}{-\hspace{-2.3ex}\int}

\begin{document}

\title{$\pi \pi$ scattering, pion form factors and chiral perturbation
  theory}


\author{Gilberto Colangelo}{
  address={Institut f\"ur Theoretische Physik der Universit\"at Bern \\
Sidlerstr. 5 3012 Bern Switzerland}
}

\begin{abstract}
I discuss recent progress in our understanding of the $\pi \pi$ scattering
amplitude at low energy thanks to the combined use of chiral perturbation
theory and dispersion relations. I also comment on the criticism
raised by Pel\'aez and Yndur\'ain on this work.
\end{abstract}

\maketitle


\section{Introduction}
In the previous conference of this series I was invited to present results
\cite{Colangelo:2002tn} concerning the experimental determination of the
$\bar q q$ condensate in the SU(2) chiral limit \cite{Colangelo:2001sp},
based on an analysis of the Brookhaven E865 data \cite{Pislak:2003sv} on
the low-energy $\pi \pi$ phase shift as extracted from $K_{e4}$ decays.
These experimental results and their analysis closed a long-standing
discussion about the size of this order parameter of chiral symmetry
breaking in QCD (cf.  \cite{Descotes-Genon:2001tn} and references therein):
the scenario in which the SU(2) $\bar q q$ condensate is unexpectedly
small, or even vanishing, though interesting, is now experimentally
excluded. Only the dependence of this condensate on the number of massless
flavours still remains an open issue (cf. \cite{Descotes-Genon:2002yv} and
references therein). In the yet earlier conference \cite{Colangelo:2000fd},
within a general discussion of recent progress in chiral perturbation
theory (CHPT), I had already presented our predictions for the two $S$-wave
scattering lengths \cite{Colangelo:2000jc}.

This work on $\pi \pi$ scattering has a few features which are worth
stressing:
\begin{itemize} 
\item the precision reached (at the level of a few percent) is quite
  unusual for hadronic physics;
\item this precision concerns a prediction -- experiments have not yet
  reached the same level of accuracy (``theory is ahead of experiment'' as
  Heiri Leutwyler puts it \cite{Leutwyler:2002hk});
\item despite a rather heavy machinery which is necessary to obtain this
  prediction, the latter does follow from QCD, and indeed the experimental
  tests tell us something about QCD, as the conclusion about the size of
  the $\bar q q$ condensate shows.
\end{itemize}
The precision obtained in our theoretical understanding of the $\pi \pi$
scattering amplitude at low energy is not only important {\em per se}, but
has also important consequences for a number of other processes. In almost
every low energy hadronic process the interaction among pions plays an
important role, and being able to treat this accurately may lead to
relevant improvements. An example of this is the anomalous magnetic
moment of the muon, where one can make good use of the accurate knowledge
of the $\pi \pi$ $P$-wave phase shift \cite{Colangelo:2003yw}.

An essential role in this improved understanding of the $\pi \pi$
scattering amplitude at low energy has been played by the combined use of
CHPT and dispersion relations. In CHPT, even after a two-loop
calculation of the $\pi \pi$ scattering amplitude, one finds out that only
close to the center of the Mandelstam triangle the series converges rather
fast, whereas at threshold the convergence is surprisingly slow: a direct
evaluation of the scattering lengths in CHPT would not have reached the
same precision level \cite{2lpipi}. On the other hand a purely dispersive
analysis of $\pi \pi$ scattering, as performed in the seventies (for a
review of this early work cf. \cite{pennington}) using Roy equations
\cite{Roy} did not lead to precise predictions either, because of lack of
information on the subtraction constants. If one uses CHPT to pin down the
latter, the scheme becomes predictive and accurate. In our work we first
had to redo the Roy equation analysis \cite{Ananthanarayan:2000ht} and then
matched the dispersive representation to the chiral one
\cite{Colangelo:2001df}.

Some of the input used in the Roy equation analysis in
\cite{Ananthanarayan:2000ht} has been criticised by Pel\'aez and Yndur\'ain
\cite{Pelaez:2003eh}, and doubts have been cast on the level of precision
reached in our analysis. This criticism has been immediately answered
\cite{Caprini:2003ta}. Also, the more recent objections raised in
\cite{Yndurain:2003vk} on the dispersive determination of the scalar radius
discussed in \cite{Colangelo:2001df} were shown to be unfounded
\cite{Ananthanarayan:2004xy}.

In the present edition of the conference a session has been devoted to a
discussion of these issues. In this contribution I will present my view on
these issues and on the ongoing discussion -- of course, the current view
of Pel\'aez and Yndur\'ain can also be found in these proceedings
\cite{Pelaez:2004xx}. Rather then concentrating on the reply to the
criticism raised by Pel\'aez and Yndur\'ain, which is rather technical and
can anyway be found in the original papers
\cite{Caprini:2003ta,Ananthanarayan:2004xy}, I will review the work we did on
the $\pi \pi$ scattering amplitude and discuss its importance also in view
of future experimental tests, as well as tests and comparisons with lattice
calculations. I will also briefly discuss the criticism and our reply, but
I wish to stress right away that the points raised in \cite{Pelaez:2003eh}
have all been answered in \cite{Caprini:2003ta}, and that the claimed
violation of a ``robust lower bound'' \cite{Yndurain:2003vk} in our
calculation of the scalar radius has been shown to be a non-issue because
this lower bound does not exist \cite{Ananthanarayan:2004xy}. The
discussion on these issues is closed. In a more recent paper
\cite{Pelaez:2004xy} Pel\'aez and Yndur\'ain claim that our representation
for the $\pi \pi$ scattering amplitude fails to satisfy some dispersion
relations. We have not yet evaluated these, and I can therefore not comment
on this claimed failure. Moreover, in their contribution to these
proceedings they criticize our choice of one of the input parameters in our
Roy analysis, the value of the $S$-wave isoscalar phase shift at $0.8$ GeV
-- I will comment on this point below.

\section{$\pi \pi$ scattering: Roy equations and CHPT}
In SU(2) CHPT the expansion parameter is $\hat m / M_\rho$ and one expects
higher order corrections to be of the order of a few percents. There are
many known examples in which this naive expectation is violated and
corrections are substantially larger. A well known example is the $\pi \pi$
$S$-wave isoscalar scattering length which has been first evaluated by
Weinberg \cite{Weinberg:1966xx} to leading order in the chiral expansion.
Numerically this gives $a_0^0(\mathrm{LO})=0.16$, but the next-to-leading
order corrections, first calculated by Gasser and Leutwyler
\cite{Gasser:1984xx} shift this value by about 25\%,
$a_0^0(\mathrm{NLO})=0.20$. The next-to-next-to-leading order corrections
have also been evaluated \cite{2lpipi} and have been found to be not yet
negligible, shifting the value by another 10\% up to
$a_0^0(\mathrm{NNLO})=0.22$. The error estimate for this quantity is not
trivial: first of all one has to determine a number of low-energy constants
(LEC) which appear in the chiral expansion of this quantity and estimate
the corresponding error. Second, one has to estimate the size of yet higher
order corrections. At first sight going below the 10\% level for the total
error appears to be difficult. The reason for the large size of the higher
order corrections for this quantity, however, is well understood and is due
to the strong interaction of the pions in the $I=0$ $S$ wave: the
perturbative expansion for these unitarity corrections converges slowly. In
a dispersive framework, on the other hand, these unitarity corrections can
be treated exactly.

If one combines the dispersive and the chiral approaches one can make a
quantum jump in accuracy: the dispersive treatment can be used to evaluate
the unitarity corrections which are problematic in the chiral expansion,
and CHPT can be used to fix the subtraction constants which represent the
only true degrees of freedom in the dispersive treatment at low energy. The
crucial point is that if one chooses the subtraction constants properly,
the chiral expansion for these does indeed follow the naive expectations
about the size of the higher order corrections. Moreover, in the low energy
region, the dispersive treatment does lead to very precise results.

I will now illustrate in some more detail this program and first discuss
the Roy equations and their numerical solution and then the matching to the
chiral representation and the numerical prediction for the scattering
lengths. 

\subsection{Roy equations}
In 1971 Roy \cite{Roy} showed that using crossing one can write a set of
dispersion relations for the $\pi \pi$ scattering amplitude which involve
only physical region singularities. When projected onto partial waves these
equations take the form of an infinite set of integral equations in which
the real part of any partial wave is given by an integral over the
imaginary parts of all partial waves in the physical region. At low energy
(say below 1 GeV) the $S$ and $P$ waves dominate, and it suffices to
consider the equations only for these lowest partial waves. For example,
the Roy equation for the $S$, $I=0$ wave reads as follows
\begin{eqnarray}
\mbox{Re}\, t_0^0(s) &=& k_0^0(s)
\hspace{-1mm} + \Pint_{4M_\pi^2}^{E_0^2} ds'
K_{0 \, 0}^{0 \, 0} (s,s')\, \mbox{Im}\, t_0^0(s')\hspace{-1mm}
+ \Pint_{4M_\pi^2}^{E_0^2} ds' K_{0 \, 1}^{0 \, 1} (s,s')\, \mbox{Im}\,
t_1^1(s') \nonumber \\  
&+& \Pint_{4M_\pi^2}^{E_0^2} ds' K_{0 \, 0}^{0 \, 2} (s,s')\, \mbox{Im}\,
t_0^2(s') + f_0^0(s) + d_0^{0}(s) \; \; ,
\end{eqnarray}
where $k_0^0$ is the contribution of the subtraction polynomial, $f_0^0$
the contribution from the intermediate energy region and $d_0^0$ the
so-called driving term containing both the contribution from the
high-energy region as well as that from the higher partial waves:
\begin{eqnarray}
k^0_0(s) &=&  a_0^0 +\frac{s-4M_\pi^2}{12 M_\pi^2} \, 
(2a_0^0-5a_0^2) \nonumber \\
f_0^0(s) &=& \sum_{I'=0}^2\sum_{\ell'=0}^1\Pint_{E_0^2}^{E_1^2} ds'  
K_{0 \ell'}^{0 I'} (s,s')\, \mbox{Im}\, t_{\ell'}^{I'}(s')  \nonumber \\
d_0^0(s) &=& \mbox{all the rest} \; \; .
\end{eqnarray}
The two energies $E_{0,1}$ were chosen in \cite{Ananthanarayan:2000ht} to
be $E_0=0.8$ GeV and $E_1=2$ GeV. With this choice of $E_0$ the solution
has been shown to be unique \cite{Gasser:1999hz}. The $f_0^0$ term is
evaluated using the imaginary parts Im$t^I_\ell$ measured in $\pi N \to \pi
\pi N$ experiments, whereas the driving terms are evaluated using
experimental information on the lowest lying resonances in $D$ and higher
partial waves, and Regge representations for the high-energy $\pi \pi$
scattering amplitude.  The kernels $K_{II'}^{\ell \ell'}(s,s')$ are all
known explicitly \cite{Ananthanarayan:2000ht}.

The upshot of the analysis in \cite{Ananthanarayan:2000ht} is that if the
input above $E_0$ is given, the solution of the equations below $E_0$ is
uniquely fixed by the two scattering lengths $a_0^0$ and $a_0^2$. Actually
only one of the two is a true free parameter if the solution has to be
physical, i.e. if it has no cusps at $E_0$. Since the input above $E_0$ is
not known with infinite accuracy, this correlation among the two input
parameters is not a line, but gets broadened into a band known as the
Universal Band. A similar correlation among the two scattering lengths is
also given by the Olsson sum rule \cite{Olsson}.

\subsection{Matching the chiral and the dispersive representation}

In the early literature on Roy equations the subtraction constants were
taken as free parameters -- in the region below the matching point $E_0$
these were the main sources of uncertainty and it was difficult to turn the
Roy equation machinery into a predictive scheme. The best use one could
make of Roy equation solutions was to analyze data in the low-energy
region, like those from $K_{e4}$ decays, in order to {\em determine} the
scattering lengths -- in much the same way as the E865 collaboration
\cite{Pislak:2003sv} has used our Roy equation solutions
\cite{Ananthanarayan:2000ht}.  The point of view has changed drastically
once it has become clear that CHPT can provide rather accurate predictions
for the scattering lengths \cite{Gasser:1984xx}.

As mentioned above, however, the scattering lengths are not the quantities
that CHPT can most accurately predict. Since the choice of the subtraction
point is arbitrary, one can exploit this freedom and subtract below
threshold, close to the center of the Mandelstam triangle, in order to be
far from the singularities that make the chiral series converge slowly.
By doing so one can optimize the accuracy of the whole scheme as
is well illustrated by the following breakdown of $a_0^0$ into
various contributions:
\begin{equation}
a^0_0  = \frac{7M_\pi^2}{32\pi F_\pi^2}\,C_0 +M_\pi^4\,\alpha_0
+O(M_\pi^8)
\end{equation}
where
\begin{equation}
C_0 = 1 + \frac{M_\pi^2}{3}\,\rs-\frac{5\, M_\pi^2 }{224 \pi^2
  F_\pi^2}\left\{\lbar_3 -\frac{563}{525} \right\}+O(M_\pi^4)
\label{eq:C0}
\end{equation}
(here given only at next-to-leading order, for simplicity) is the part of
the scattering length which only depends on the quark masses, whereas
$\alpha_0$ is the part which is due to the momentum dependent part of the
amplitude, here evaluated at threshold. 
\begin{figure}[thb]
 \includegraphics[width=0.9 \textwidth]{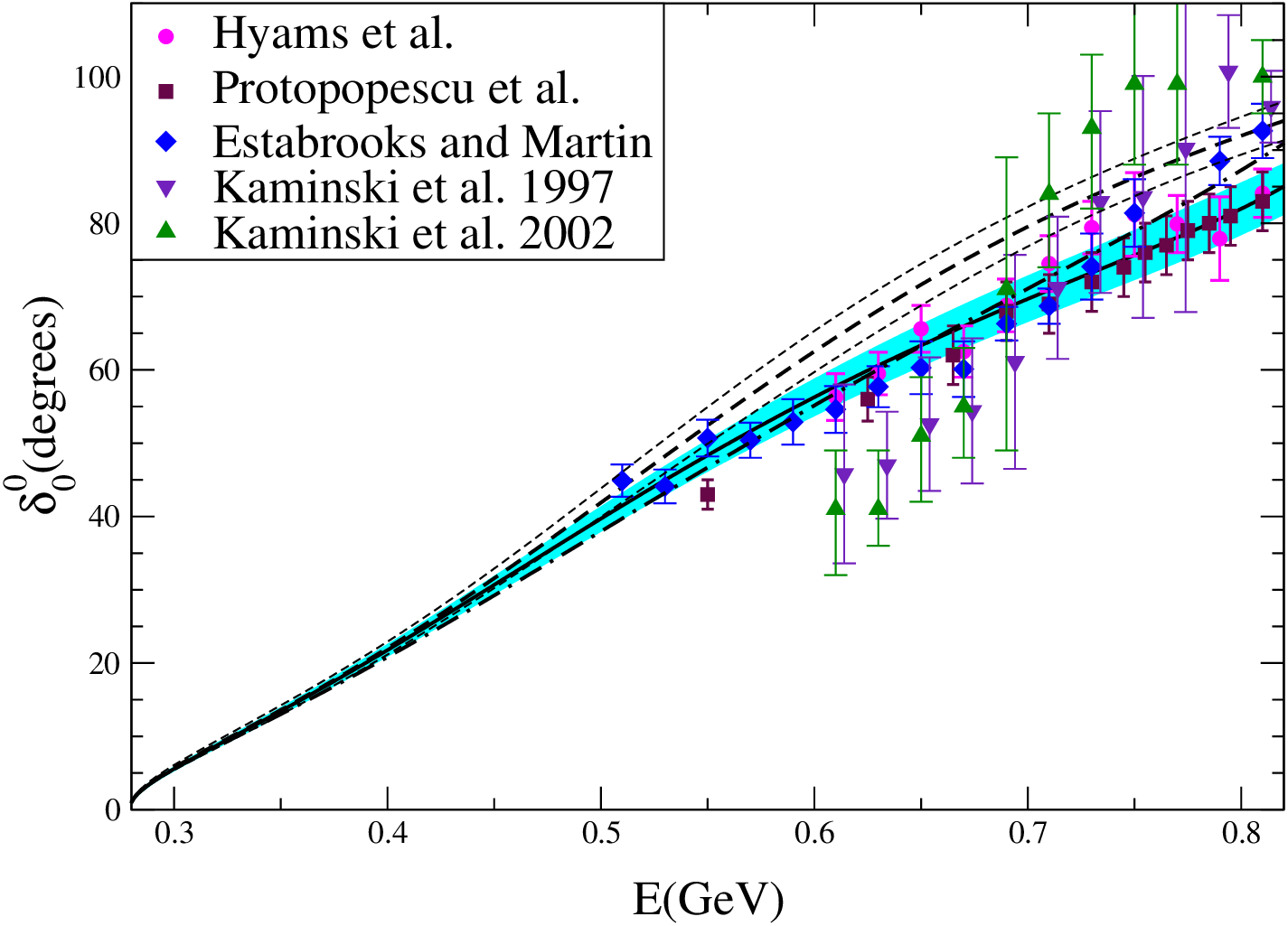}
 \caption{\label{fig:Swave} Phase shift for the $I=0$ $S$ wave. The shaded
   band is the result of the analysis in \cite{Colangelo:2001df} and the
   solid line is the Roy equation solution obtained with the input above
   $1.4$ GeV proposed in \cite{Pelaez:2003eh}. The dashed curve and the
   corresponding uncertainty band are the ``tentative alternate solution''
   proposed in \cite{Pelaez:2003eh}. The dotdashed curve is the Roy
   solution fit to the Kaminski et al. 1997 \cite{KLR} data obtained in
   \cite{Kaminski:2002pe}. The other data sets shown are from
   Refs.~\cite{hyams, proto, EM}}
\end{figure}
The constant $C_0$ is to be chosen
as subtraction constant in the dispersive treatment, because for this the
chiral expansion converges fast: the numerical evaluation at the two-loop
level gives
\begin{equation}
C_0=1.096 \pm 0.021
\end{equation}
amounting to a 10\% shift evaluated with 20\% of relative uncertainty. This
correction shifts $a_0^0$ from $0.16$ to $0.17$. The bulk of the
correction, however, comes from the momentum dependent part of the
amplitude, the $\alpha_0$ term which can be accurately evaluated through a
dispersive integral. The final result for both $S$-wave scattering lengths
reads
\begin{equation}
a_0^0=0.220 \pm 0.005 \qquad a_0^2=-0.0444 \pm 0.0010 \; \; ,
\label{eq:aI0}
\end{equation}
with an accuracy at the level of a few percent for both. Notice that the
situation is completely different for $a^2_0$: here the tree level result
is $-0.0454$ and the momentum-dependent part of the correction is not
particularly large. As is well known the $\pi \pi$ interaction in the $I=2$
channel is weak.

\begin{figure}[thb]
 \includegraphics[width=9cm]{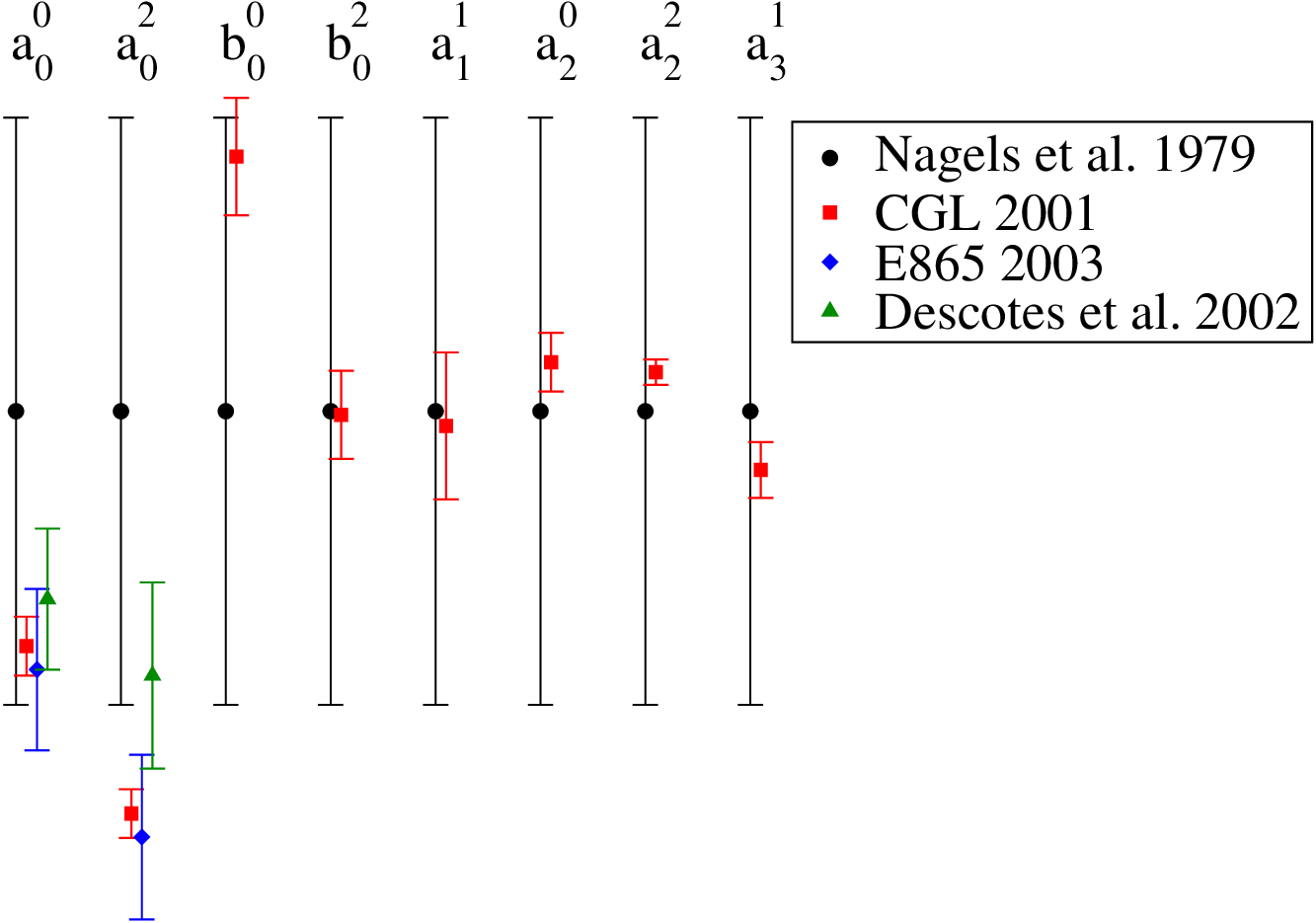}
 \caption{\label{fig:threshold} Accuracy improvement in the evaluation of
   various threshold parameters due to the use of CHPT for the subtraction
   constants.}
\end{figure}
If one uses the dispersive representation for the $\pi \pi$ scattering
amplitude, this high level of accuracy obtained for the scattering lengths
is reflected in the whole energy region below $E_0$: having fixed the
scattering lengths all other sources of noise generate remarkably little
uncertainty as illustrated for the case of the $S0$ wave in
Fig.~\ref{fig:Swave}.  Notice that the band is a prediction which relies on
experimental input only at $0.8$ GeV and above -- it is {\em not} a fit to
any of the data sets shown in the same plot.  The plot contains various
data sets as well as the ``tentative alternate solution'' proposed by
Pel\'aez and Yndur\'ain in \cite{Pelaez:2003eh} and the Roy solution fit of
Kaminski et al.  \cite{Kaminski:2002pe}. These will be commented upon in
the next section.  That CHPT is mostly responsible for the improvement in
the accuracy in the Roy treatment is well illustrated in
Fig.~\ref{fig:threshold}, where the values for a number of threshold
parameters obtained from Roy equation analyses as reported in the
compilation of data \cite{Nagels} are given in arbitrary units chosen such
that all errors are normalized to the same size.  The errors obtained after
matching the dispersive and the chiral representation are about an order of
magnitude smaller.

\section{The criticism of Pel\'aez and Yndur\'ain}
In \cite{Pelaez:2003eh} Pel\'aez and Yndur\'ain have criticized our work
\cite{Ananthanarayan:2000ht,Colangelo:2001df} and made the following
claims: that the input used in the Roy equation analysis above 1.4 GeV was
incorrect and that as a consequence the solutions we obtained below 0.8 GeV
were also incorrect. The claim about the incorrectness of the input has two
aspects: between 1.4 GeV and 2 GeV we used experimental information and
they claim that this is unreliable -- above 2 GeV, where we relied on a
Regge representation, they claimed that the one we used is not orthodox
because it does not respect factorization. According to the latter
property, the residues of the Regge poles which appear in the $\pi \pi$
scattering amplitude must be given by the square of the residue for $\pi N$
divided by the $NN$ residue.  As explicitly stated in
\cite{Ananthanarayan:2000ht} the Regge representation we used served the
purpose of giving us a fair account of the contributions to the dispersive
integrals from the regions between 2 and 3 GeV -- the contributions from
yet higher energies are negligible because the Roy equations are twice
subtracted. In fact even the contributions from the region above 1.4 GeV
are rather small and play a minor role in the Roy equations. For this
reason we have not made our own analysis of this part of the input and took
what was available in the literature.

The observation that contributions from above 1.4 GeV do not matter much
for the Roy solutions below 0.8 GeV appears to be in contradiction with the
second claim made by Pel\'aez and Yndur\'ain, namely that the solutions we
obtained were ``spurious'' because of the incorrect input used. It is
important to stress here that Pel\'aez and Yndur\'ain made this claim
without supporting it with a calculation, but only with indirect
arguments. I will come back to these indirect arguments later but I first
must say that the Roy equation solutions for the input proposed by Pel\'aez
and Yndur\'ain as the correct one have been calculated in
\cite{Caprini:2003ta}. The outcome is the solid line in
Fig.~\ref{fig:Swave} and is indistinguishable from the solution obtained
with the input originally used in \cite{Ananthanarayan:2000ht}. The
calculation shows that the second claim of Pel\'aez and Yndur\'ain is
wrong.

This takes us back to the indirect arguments they had used to support
their claim. They gave three arguments:
\begin{enumerate}
\item a mismatch in the Olsson sum rule;
\item a discrepancy among two different determinations of the $P$-wave
  effective range;
\item a discrepancy among two different evaluations of the $D$ and $F$ wave
  threshold parameters.
\end{enumerate}
In \cite{Caprini:2003ta} we have discussed in detail all these indirect
arguments and shown that either the discrepancy is not there (as in the
case of the $P$ wave effective range) or that the conclusion that the
discrepancy can be cured by changing the Roy solution below 0.8 GeV is
incorrect. The interested reader is referred to \cite{Caprini:2003ta} for a
detailed discussion of all these points.

As explicitly stated also in \cite{Caprini:2003ta}, a better input than the
Regge representation that we have used in \cite{Ananthanarayan:2000ht} is
certainly possible and can obtained with a thorough analysis of all the
available information (like high-energy total cross section data, sum rules
etc.) \cite{Caprini-WIP}. In this perspective, also the data on total $\pi
\pi$ cross sections which Pel\'aez and Yndur\'ain have pointed out
\cite{Pelaez:2004xx} are useful information which has to be taken into
account -- the representation used in \cite{Ananthanarayan:2000ht} does not
describe these very well and can be improved. It is however a fact that the
influence of improvements in the high-energy input on the scattering
lengths and the whole low-energy scattering amplitude will be negligible.
Such improvements may be of interest in applications which rely on the $\pi
\pi$ scattering amplitude close to $1$ GeV, like the dispersive
representation of hadronic contributions to $a_\mu$ which relies on the $P$
wave phase shifts \cite{Colangelo:2003yw}.

\subsection{The input phase $\delta_0^0(0.8 \mbox{GeV})$}
In the most recent papers Pel\'aez and Yndur\'ain have moved on to discuss
other points and to raise further criticism to our analysis. In particular
they have criticized our value and error for $\delta_0^0(E_0)$ which, as
discussed at length in \cite{Ananthanarayan:2000ht}, is one of the most
important input parameters in the Roy analysis.
\begin{figure}[th]
 \includegraphics[width=9.5cm]{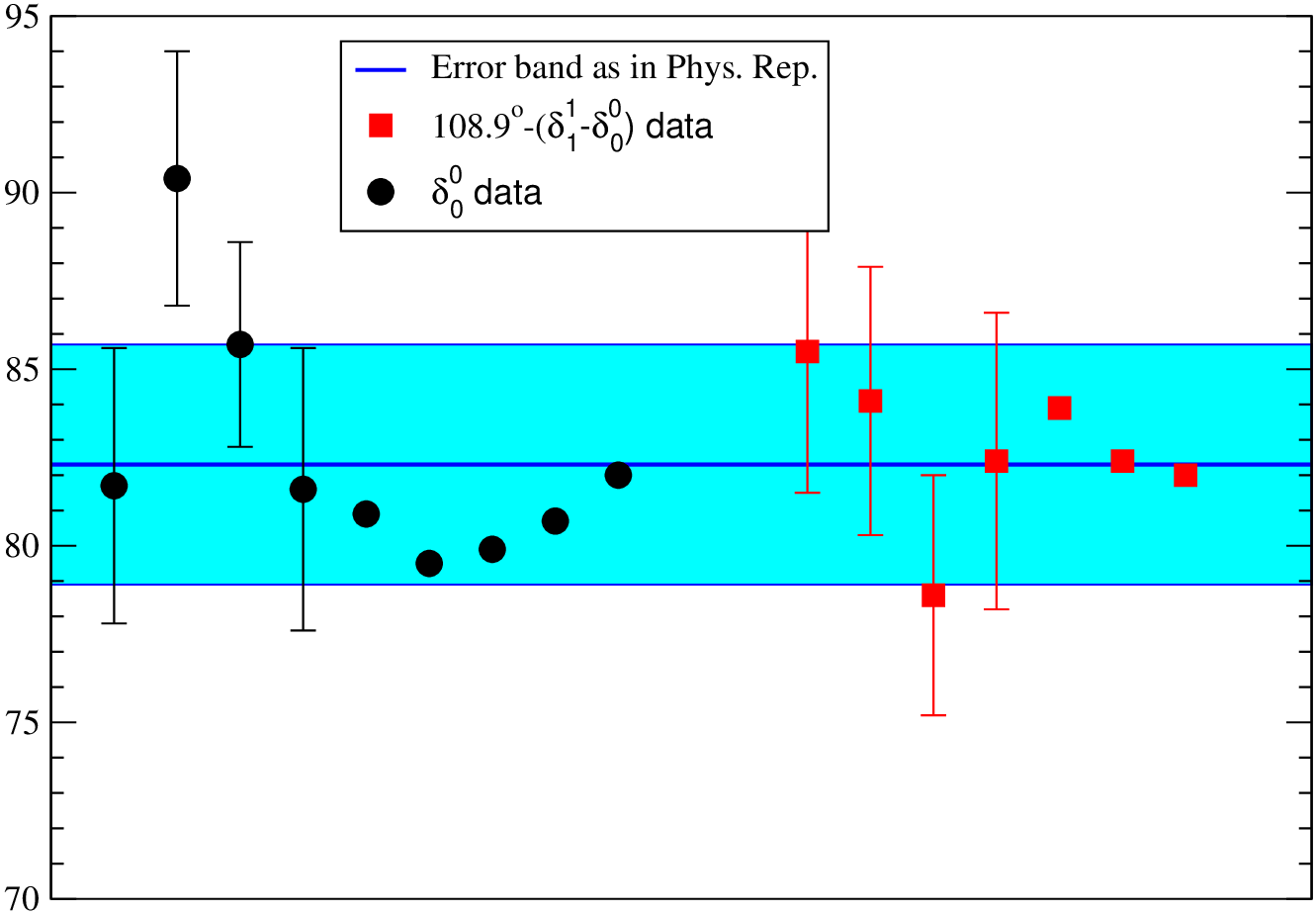}
 \caption{\label{fig:d00d11} Graphical representation of the data on
   $\delta_0^0(0.8 \mbox{GeV})$ as given in Table 2, p. 226 of
   Ref.~\cite{Ananthanarayan:2000ht}: on the left-hand side of the plot the
   plain data for $\delta_0^0(0.8 \mbox{GeV})$ are shown. On the right-hand
   side those for the difference $\delta_0^0(0.8 \mbox{GeV})-\delta_1^1(0.8
   \mbox{GeV})$ shifted by $108.9^\circ$, the value of the $P$ wave phase
   shift as extracted from form factor data.}
\end{figure}
In this paper we had observed that if one looks at the data on this
particular wave it is difficult to draw any conclusion because different
data sets (in fact different analyses of the same $\pi N \to \pi \pi N$
scattering data) are mutually incompatible.  The situation improves
dramatically if one looks at the difference
$\delta_1^1(E_0)-\delta_0^0(E_0)$, for which different data sets give a
coherent picture. The fact that the $\delta_1^1$ phase is now known much
better thanks to the data on the vector form factor then leads to a rather
good determination of $\delta_0^0(E_0)$ which we estimated to be
$82.3^\circ \pm 3.4^\circ$. This is illustrated in Fig.~\ref{fig:d00d11}
which compares the direct determinations of $\delta_0^0(E_0)$ to those that
exploit the phase difference.  In their recent discussions Pel\'aez and
Yndur\'ain have given particular emphasis to the recent analyses by
Kaminski et al. \cite{KLR}. This is indeed important new work on $\pi \pi$
scattering because it analyzes a large body of polarized $\pi N \to \pi \pi
N$ data, which are certainly very interesting and useful. The data of these
two analyses are shown in Fig.~\ref{fig:Swave}: around 0.8 GeV these data
are substantially higher than those of Hyams et al. or Protopopescu et al.
and our band. In that region the ``tentative alternate solution'' proposed
by Pel\'aez and Yndur\'ain in \cite{Pelaez:2003eh} is in better agreement
with them. Below 0.7 GeV, however, the Kaminski et al. data become lower
than the other data sets and also lower than the two bands shown: even the
``tentative alternate solution'' is now in flat disagreement with these
data. The problem is seen also in a recent paper by Kaminski et al.
\cite{Kaminski:2002pe} where they solve Roy equations and fit the data with
the solutions: the overall fit is not particularly good precisely because
this peculiar shape of the data in this energy region (low between 0.6 and
0.7 GeV, rising steeply at 0.7 GeV and then high until 0.8 GeV) cannot be
followed by Roy equation solutions. This best fit to their data is shown in
Fig.~\ref{fig:Swave} as dotdashed curve, and is evidently a trade-off
between the high and the low data. This curve lies almost everywhere inside
our Roy solution band and has $\delta_0^0(0.8 \mbox{GeV})=87^\circ$, about
1.5 $\sigma$ higher than the range we had used as input.

This most recent analysis does not clarify the experimental situation
concerning the $S0$ wave. Since no information is provided on the phase
difference $\delta_1^1-\delta_0^0$ we could not make use of these data when
we fixed the input for the Roy equations. Moreover the results of the Roy
equation analysis of Kaminski et al. \cite{Kaminski:2002pe} we just
discussed show that there is no reason to modify our central value for
$\delta_0^0(0.8 \mbox{GeV})$ or to stretch its error.

\subsection{Scalar radius}
Another point on which criticism has been raised against our analysis
concerns the scalar radius $\rs$ which appears in the chiral expansion of
both $S$-wave scattering lengths, cf. Eq.~(\ref{eq:C0}). The input value we
used \cite{Colangelo:2001df}:
\begin{equation}
\rs=0.61 \pm 0.04 \mathrm{fm}^2
\label{eq:r2}
\end{equation}
had been determined through a dispersive analysis of the scalar form
factor, following \cite{Donoghue:1990xh} after updating the $\pi \pi$ phase
shifts which are used as input. Yndur\'ain has recently claimed that the
outcome of this calculation violates a ``robust lower bound'' he derives
\cite{Yndurain:2003vk}. The same bound is violated also by other
calculations of the same quantity. One of them, performed by Moussallam
\cite{Moussallam:1999}, makes a thorough analysis of the different
phenomenological inputs available in the literature and comes to a
conclusion which is in perfect agreement with Eq.~(\ref{eq:r2}): the values
he finds for the radius lie between $0.58$ and $0.65\,\mbox{fm}^2$.

This apparent puzzle has been recently clarified in
\cite{Ananthanarayan:2004xy}: the ``robust lower bound'' does not exist
because its derivation relies on an untenable assumption, namely that the
phase of the scalar form factor in the region above $1.1$ GeV must be close
to the $S0$ phase shift. The correct conclusion is that above $1.1$ GeV
where the elasticity is again close to one, the difference between the
phase of the scalar form factor and the phase shift has to be close to a
multiple of $\pi$: the various available calculations all agree that this
difference is close to $\pi$ rather than to zero. Also this criticism is
unjustified.

\section{Outlook}
The predictions for the $S$ wave $\pi \pi$ scattering lengths
(\ref{eq:aI0}) discussed here are of unusual precision in hadronic
physics. They are derived under the assumption that the quark condensate is
the leading order parameter of the spontaneous symmetry breaking in
QCD. Experimental tests have confirmed this hypothesis and have put the
standard picture of the QCD vacuum on a solid experimental basis. The
predictions are however still more precise than the experimental
measurements of the same quantities, and there is still room for a more
stringent test of QCD at low energy.  Experiments aiming at performing
these tests are currently underway: the DIRAC experiment at CERN
\cite{DIRAC} is measuring the lifetime of pionic atoms which is
proportional to the square of the difference of the two scattering lengths
and plans to reach a 5\% precision in the measurement of this
difference. The NA48II experiment \cite{NA48II} will gather twice the
statistics of $K_{e4}$ decays of the E865 experiment, thus reaching an
improvement of a factor $\sqrt{3}$ in the final uncertainty. The phase
shift difference measured in these decays is mostly sensitive to $a_0^0$.
The prospects for precision low-energy hadronic physics are at the moment
particularly good. We look forward to the experimental results.

\begin{theacknowledgments}
It is a pleasure to thank the organizers for the invitation and the perfect
organization of the conference in a wonderful environment. I also thank
B.~Ananthanarayan, I.~Caprini, J.~Gasser and H.~Leutwyler for the
longstanding and very pleasant collaboration on the topics discussed here
and for useful comments and suggestions on the manuscript.
\end{theacknowledgments}

\end{document}